\documentclass[runningheads]{llncs}
\usepackage[utf8]{inputenc}
\usepackage{url}
\usepackage{graphicx}
\usepackage [english]{babel}
\usepackage [autostyle, english = american]{csquotes}
\MakeOuterQuote{"}

\title{Understanding scholarly Natural Language Processing system diagrams through application of the Richards-Engelhardt framework}
\titlerunning{Understanding scholarly NLP system diagrams}
\author{Guy Clarke Marshall, Caroline Jay, and Andr\'e Freitas}
\institute{University of Manchester}
\date{January 2020}
\begin{document}

\maketitle

\begin{abstract}
We utilise Richards-Engelhardt framework as a tool for understanding Natural Language Processing systems diagrams. Through four examples from scholarly proceedings, we find that the application of the framework to this ecological and complex domain is effective for reflecting on these diagrams. We argue for vocabulary to describe multiple-codings, semiotic variability, and inconsistency or misuse of visual encoding principles in diagrams. Further, for application to scholarly Natural Language Processing systems, and perhaps systems diagrams more broadly, we propose the addition of "Grouping by Object" as a new visual encoding principle, and "Emphasising" as a new visual encoding type. 

\end{abstract}
\section{Introduction}

The framework proposed by Engelhardt and Richards \cite{engelhardt2018framework} is designed to be applicable to "all types of visualizations". Their framework provides a method and vocabulary for analysing diagrams and diagram components, which they have tested on a large variety of visualization types. It appears the framework has not yet been applied to system diagrams. We aim to advance the understanding of the heterogeneous diagrammatic representations of Natural Language Processing (NLP) systems found in scholarly conference proceedings by applying this framework. The scholarly NLP systems diagrams domain has been chosen due to its unusual heterogeneity, and its importance and prevalence in communicating about NLP systems research. Further, we propose modest clarifications or extensions to the framework in order to fit this domain, which may benefit the wider domain of system diagrams. Whilst Engelhardt and Richards tested their framework on a wide variety of information visualisation resources they did not examine (a) systems diagrams, or (b) "ecological" diagram examples not conforming to an established standard or grammar. Our method is to:
\begin{enumerate}
    \item Manually select a variety of natural language processing system diagrams in recent conference proceedings
    \item Apply the Richards-Engelhardt framework in order to analyse these examples
    \item Discuss and make suggestions to improve the applicability of Richards-Engelhardt framework in this domain
	\item Discuss implications of this reflection for the scholarly NLP community
\end{enumerate}

In the course of this analysis, we necessarily comment on some existing diagrams. Our goal is not to criticise the authors of these diagrams, but rather to support the community in creating effective diagrams. Our contribution is to:
\begin{itemize}
	\item[\textbullet] Conduct diagram analysis on scholarly NLP system diagrams
	\item[\textbullet] Qualitatively evaluate Richards-Engelhardt framework as a tool for reflecting on scholarly NLP systems diagrams
	\item[\textbullet] Suggest further requirements for the framework in this domain, including emphasis, language, graphical object schematicity, and semiotic considerations
\end{itemize}

\section{NLP Diagrams}
\subsection{NLP Systems}
Natural Language Processing is a discipline within Computer Science, and is concerned with creating systems that solve tasks relating to Natural Language interpretation. NLP systems take a text input, go through data manipulation steps, and create an output that is usually a classification or a prediction, such as what the next word in a sequence is likely to be. The state-of-the-art systems are technically complex, requiring application of mathematical and algorithmic techniques. These NLP systems are often described through diagrams. We have chosen to examine \textit{scholarly} NLP system diagrams found in conference proceedings, as a sub-domain of NLP systems diagrams with a well-defined scope. 
\subsection{NLP System Components}
Modern NLP systems are often based on neural networks, and it is these systems we focus on. A neural network takes an input (in NLP, text), and then processes this via a series of \emph{layers}, to arrive at an output (classification/prediction). Within each layer are a number of \emph{nodes} that hold information and transmit signals to nodes in other layers. Specific mathematical functions or operations are also used in these systems, such as sigmoid, concatenate, softmax, max pooling, and loss. The \emph{system architecture} describes the way in which the components are arranged. Different architectures are used for different types of activities. For example Convolutional Neural Networks (CNN), inspired by the human visual system, are commonly used for processing images. Long Short Term Memory networks (LSTM), a type of Recurrent Neural Network (RNN) which are designed for processing sequences, are often used for text.

These neural networks "learn" a function, but have to be trained to do so. Training consists of providing inputs and expected outputs, allowing the system to develop an understanding of how an input should be interpreted. The system is then tested with unseen inputs, to see if it is able to handle these correctly. System diagrams almost always depict the training process. A more detailed introduction to LSTM architectures, including schematics, is provided by Olah~\cite{olah2015understanding}.

\subsection{NLP scholarly documents}
The representation of these systems in conference proceedings is done in natural language, in pseudocode, and in diagrams which often appear to describe the system beyond the neural network itself, including inputs, outputs, and the relationship between components. Code may be shared as a supporting artifact external to the formal proceedings. System outputs are often shared as tables, charts, and metrics. 
In order to consider diagrams separately from the text, we focus on \emph{self-contained diagrams} which were deemed to meaningfully exist with assumed knowledge but without reference to other resources.
\subsection{NLP scholarly diagrams}
\subsubsection{Overview}
In this domain, representation can be challenging and complex. As noted by Mahoney \cite{mahoney1985diagrams} in engineering, there are signification issues associated with representing how things work rather than what they are:
\begin{quote}
"But to show what machines do or how they are assembled is one thing; to show how they work is quite another. However accurately and fully a complex mechanism may be portrayed, an understanding of its operation as a whole rests ultimately on familiarity with the operations of its basic components. Treatises of the genre under discussion took that familiarity for granted. Their authors could not do otherwise, given the nature of their medium. A picture of a windlass, or of a system of pulleys, cannot in and of itself set forth the laws that define the device's mechanical advantage. A drawing of a closed tube standing in a pool of water and having a piston with a valve that opens in one direction only will still not explain a water pump until the readers know the laws (or at least the rules of thumb) that link the reduction of air pressure to the rise in the head of a column of liquid. Readers must bring knowledge or experience of such matters to the illustrations in order then to appreciate or profit from the ingenuity with which the basic machines are combined or adapted to particular circumstances." \cite[p.201]{mahoney1985diagrams}
\end{quote}

Diagrammatic representation and usage in this context lacks scientific investigation. Indeed, scholarly diagrams more broadly have been neglected by scholarly enquiry, despite their potential to enhance the communicative effectiveness of research outputs. An exception to this is within automated computer vision, where there is increasing interest in classifying scholarly figures \cite{siegel2016figureseer}, and particularly scholarly charts of experimental results \cite{ray2015architecture,savva2011revision}. 

\subsubsection{Analysing scholarly diagrams in Computer Science}
In these diagrams, apart from context related aspects such as diagram usage, two aspects of research interest are (a) what is represented and (b) how it is represented. For examples, see Figs. \ref{fig:layercentric}, \ref{fig:datacentric}, \ref{fig:chess} and \ref{fig:lebanoff}.

We are not aware of research applying existing diagram analysis frameworks to scholarly communication. One such framework is "Physics of Notations", which is highly cited for designing visual notations \cite{moody2009physics}. This could potentially be re-purposed as an analytic lens to describe scholarly-NLP-domain-specific phenomena, as it has been for investigating the impact of the addition of colour in UML Activity Diagrams found in Software Engineering \cite{gopalakrishnan2010adapting}. However, Physics of Notations is fairly abstract, with categories such as "include explicit mechanisms for dealing with complexity". We choose to progress with the Richards-Engelhardt framework due to its concreteness, its task-independence, and its potential for conducting systematic analysis. We also consider further development of the framework as part of the analysis process. 


\section{Application of the Richards-Engelhardt framework to the NLP systems diagram domain}
\subsection{Methodology}
The Richards-Engelhardt framework consists of three modes, five types, and 15 principles of visual encoding, and provides a systematic method for analysing diagrams and visualisations. In its present form the framework focuses on the principles of visual encoding, which we also focus on in our analysis. Within the visualisation design space, system diagrams in NLP could broadly be considered flow diagrams, though they do not conform to a consistent or standard form. Examining self-contained diagrams found at ACL 2019, a top NLP conference, we found a visually and semantically heterogeneous set of diagrams across the proceedings. From this set, we selected four diagrams to investigate further. They were chosen to be distinct in terms of mode of visual encoding, and the information they convey.  

\subsection{Aspects of visual representation in NLP systems diagrams} \label{subsection:VR}
We apply the Richards-Engelhardt framework in the NLP system diagram domain. Our application area falls within Engelhardt and Richard's scope of 2D static representations, though this is a novel and more complex area of application, as we shall demonstrate. We proceed to consider the three representational modes, followed by a general discussion of visual encoding types in NLP, and discuss in depth visual encoding principles of four examples from this domain. 


\paragraph{Mode of correspondence} In NLP systems diagrams, and indeed all diagrams of software, all visual encoding is necessarily \textbf{non-literal}, in that the relationship is conceptual and metaphorical. This is not trivial, since the functions (or code) that the system comprises have specific virtual inputs and outputs. It is also the case that software architecture diagrams often do not represent how the code is written, but rather how the system works, being more about signifying function than about representing code in a "realistic" sense. 

\begin{figure}[htbp]
    \centering
    \includegraphics[scale=0.8]{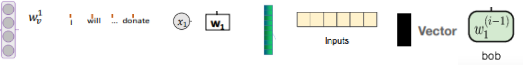}
    \caption{Nine different graphical components representing vectors found at ACL 2019. Left to right: Circles: [Zang et al, ACL 2019]. Vector: [Lee et al, ACL 2019]. Example: [Wang et al, ACL 2019]. $x_1$ circle: [Ma et al ACL 2019]. $w_1$ square: [Sarkar et al, ACL 2019]. Stacked squares: [Mai et al, ACL 2019]. Horizontal squares: [Mihaylova and Martins, ACL 2019]. Rectangle: [Barezi and Fung, ACL 2019]. Curved corners, index plus example [Zhang et al, ACL 2019]} 
    \label{fig:vectors}
\end{figure}

\paragraph{Mode of depiction} It would be challenging to describe NLP system diagrams on a realistic/precise to schematic dimension. To highlight this, Fig. \ref{fig:vectors} shows some of the different graphical components used in visual representations of vectors (which in this context are indexable lists of elements), a specific and prevalent aspect of NLP systems. Some are more "schematic" that others, in the sense that they use less ink to signify the same thing (a vector of unknown dimension). There can also be visual concreteness (such as four circles in a vector, seen left of Fig. \ref{fig:vectors}), which does not necessarily signify precision or semantic relevance (in the example, there may not be four elements). In this domain, perhaps realistic and precise are fundamentally different: Often where we find a precise depiction, it does not depict precision.

Further specific discussion of depiction could include whether graphical components should include "mathematical symbols". There are other standard recurrent symbols, such as ellipses or domain specific iconography, which could be captured by sub-categories of "depicting".

\paragraph{Mode of visual encoding} For "depicting", the NLP system diagram, as a technical illustration, is picturing (rather than mapping), and this must be the key visual encoding depiction principle. Again due to the lack of physical location, we are not in the "mapping" category. With a more metaphorical, less physically grounded, but equally rational definition, perhaps an NLP system diagram could be considered a map of the functionality (cf. \cite{dodge2001atlas,peterson2002maps}). In other domains, a "process map" would not fit naturally within the definition of either mapping or depicting categories.

The above domain-level features are applicable throughout diagrams used in this domain. We proceed to use the guidance in Fig. 1 of \cite{engelhardt2018framework} to apply the "types of visual encoding" to our domain. A collection of diagrams can be found at \url{aidiagrams.com/resources}, on which this more general commentary is based. Examining this larger sample of over 150 scholarly NLP system diagrams allows for an awareness of the heterogeneity of the domain.
\subsubsection{Visual encoding types}

\paragraph{Scaling} This is variable between and within diagrams. Within a diagram, size can be used to represent dimensional differences, often indicating a binary "bigger or smaller than", rather than precise scaling (see Fig. \ref{fig:datacentric}). 

\paragraph{Ordering} Often these diagrams are (broadly) read linearly left-right, in chronological order of data processing at training-time. The diagrams are not usually of the system at run-time. Some information important for the creation and operation of the system, including chronological information such as "parameter training process", intervals, epochs, and updating of parameters are often omitted. 

\paragraph{Grouping} Varied, often multiple different encodings are used within each diagram. Often, multiple visual encodings are applied to perform the same grouping (such as nesting, colour, proximity, alignment, and boundary, plus a label and a caption, as shown in Fig. \ref{fig:layercentric}).

\paragraph{Linking} This is perhaps the most important part of the NLP system diagrams, since along with components themselves, the relationship between components determines the architecture. Arrows and nesting are often used. 

It remains to consider visual encoding principles, which we will through analysis of four example diagrams extracted from scholarly conference proceedings.

\begin{figure}[htbp]
    \centering
    \includegraphics[scale=0.6]{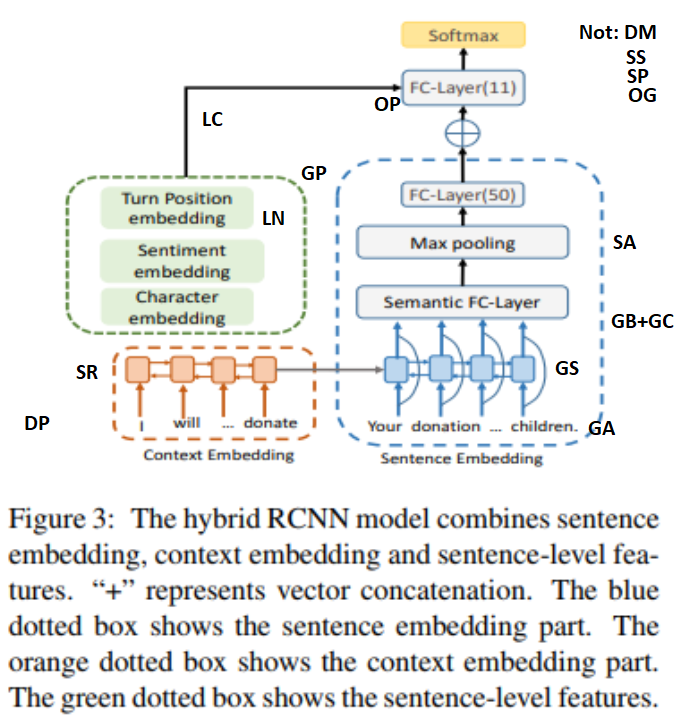}
    \caption{A layer-centric example [Wang et al, ACL 2019], with labels from the framework overlaid}
    \label{fig:layercentric}
\end{figure}

\subsection{Example diagram prioritising "layers": Fig. \ref{fig:layercentric}}
The diagram in Fig. \ref{fig:layercentric} prioritises depiction of the layers of the neural network architecture, including labelled FC-Layers (Fully Connected layers, used for classification).

Colour and boundary are consistently employed simultaneously for grouping, together with labels-attached-by-proximity for the red and blue sub-figures. Alignment groups the right section as the main components of the system (comprised of the FC-Layers). The green component is grouped by proximity to the other "early-in-the-workflow" item of context embedding. 

This diagram contains 23 words, and of those five (over 20\%) are "embedding". That does not include the caption, which (quite unusually) reemphasises the labels by describing the visual techniques of the diagram, and clarifying a symbol which is used in an unconventional way (and does not match the same symbol in the diagram, "$+$" is neither visually nor semantically equivalent to "$\bigoplus$"). It would be possible to rearrange this diagram with less natural language and the same semantic content, for example by making use of the Nesting visual encoding principle. This is an example of how an awareness of Richards-Engelhardt framework has the potential to help diagram authors with novel representations. Engelhardt and Richards have already assessed hundreds of diagram types which work well with their framework, so we focus our narrative on diagrams for which it does not work so well, such as where the author makes unconventional and internally inconsistent visual encoding choices, where communicative language is mixed, or where a visual encoding principle is used for emphasis.

\begin{figure}[htbp]
    \centering
    \includegraphics[scale=0.5]{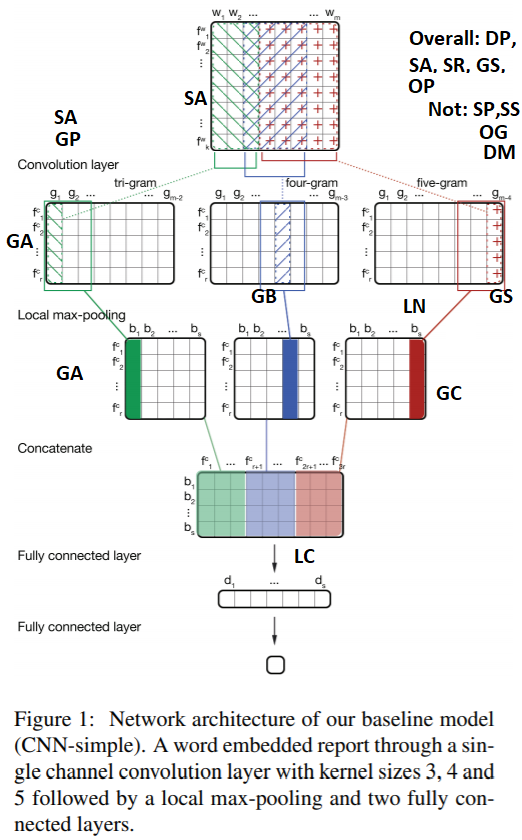}
    \caption{A data-centric example [Dereli and Saraclar, ACL 2019], with labels from the framework overlaid}
    \label{fig:datacentric}
\end{figure}

\subsection{Example diagram prioritising data manipulation: Fig. \ref{fig:datacentric}}
Fig. \ref{fig:datacentric} shows a data centric depiction of a neural network system. The dominating feature (in terms of both ink and semantic content) are the grids which represent data flows. Note that two visual techniques (gradient and scaling) are used but do not mean what the framework suggests. Colour gradient is not ordering (OG), but rather is descriptive of dilution. Scaling by proportion (SP) occurs in the concatenate layer, but is not what is meant by the use of Scaling in the local max-pooling or convolutional layer. This diagram also showcases inconsistent dual-coding, highlighting the arbitrary dimensions used throughout this domain: The sizing is not consistent with the dimensional labeling within each layer, which implies they are different widths (such as the convolutional layer, which has variable dimensions implied by the mathematical notation $g_{m-2}$ and $g_{m-4}$). The framework does not allow us to discuss author-assumed conventions (e.g. ellipsis, acronyms, subscript notation). Also in this diagram we have arrows for the last two layers but not elsewhere, and the linking is semantically different. With an implicit axis, it is understandable that the authors did not include explicit arrows as is usual for system flow. In Fig. \ref{fig:datacentric}, we found the caption to be very helpful in interpreting the overall diagram. This motivates the inclusion of the caption as part of the scope of the diagram, though they do not have the spatial relation required of a diagram and were not discussed in Engelhardt and Richards' examples.

\begin{figure}[htbp]
    \centering
    \includegraphics[scale=0.35]{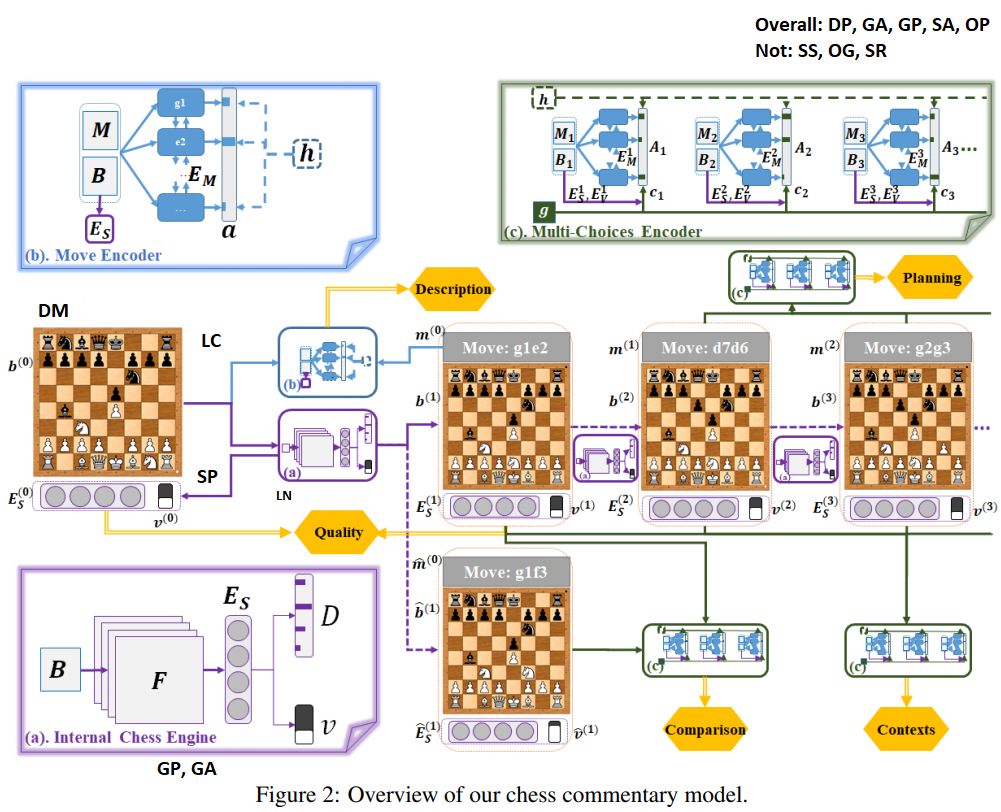}
    \caption{An example with sub-figures [Zang et al, ACL 2019], similar in size to the original, with labels from the framework overlaid}
    \label{fig:chess}
\end{figure}

\subsection{Example diagram with picturing and sub-figures: Fig. \ref{fig:chess}}
We chose Fig. \ref{fig:chess} due to its visual appeal, range of visualisation choices, and clear sub-figures allowing for easy reference without needing to overlay additional information. With the depiction of chessboards, this is depiction by mapping, as well as the dominant depiction by picturing.
Fig. \ref{fig:chess} has many different grouping by boundary mechanisms, where the visual objects used to establish the boundary are also grouped by colour and shape (wide brush with rounded corners, thin brush with rounded corners, thick brush with "folded" corner). Grouping is done by position and alignment, particularly within the labelled sub-figures (a), (b) and (c). Additionally, proximity is used to conceptually attach the vectors (collection of circles) to the chess boards, certainly in the leftmost instance. The other instances have a dotted yellow boundary (if you look very closely!). It is not clear why there is this inconsistency.

There are some visual inconsistencies which do not appear to represent meaning. Indefinite continuity of the process, in this case, is represented by one ellipsis, that is aligned only with the rightmost chessboard. Other lines simply stop. It would seem better to have the ellipsis positioned further right.
Ellipses are also used within sub-figure (b) to indicate arbitrary dimension, which is consistent with the other use. However, in sub-figure (a), the 4 circles of vector "$E_S$", the implied four overlaid squares of "$F$", and the size and number of bars in "$D$", as well as the stacked bar charts $v$, are precise and arbitrary. In the text it is made clear that the number of feature planes "$F$" is 20. It is not clear why different representational choices have been made for the same concept.
The yellow arrows are double-barred, further distinguishing them from the other coloured arrows. The rightmost two arrows in (a) have a lower weight than any other arrow, presumably an oversight. The overall ordering of the diagram is slightly unclear, with no single flow directed by linking or position. This is further complicated by the sub-figures. 
The framework does not encompass languages involved, which can be argued to be a specific visual encoding, or at very least a special type of graphical component. With example usage from Fig. \ref{fig:chess}, we have: English ("Quality"), mathematical ("$\hat{m}^{(0)}$"), Smith chess notation ("g1e2"). This would be useful, to facilitate comparison between diagrams, as a diagram with a different language has a different set of implicit assumptions made about the prior knowledge of the reader. 

The size of the chessboards and the iconic sub-figures used within the main diagram (blue and purple) vary, presumably for pragmatic size-constraint reasons. It is also potentially distracting for readers that the scaling of the (blue, purple and green) icons preserved the size of the arrowheads (only), and that purple introduces curvature to arrows.


The framework has allowed a discussion of this multifaceted diagram. Perhaps a framework omission in describing Fig. \ref{fig:chess} is that this is semiotically different to Figs. \ref{fig:layercentric} and \ref{fig:datacentric}, in that the diagram shows a specific example (of the processing of a single chess move) which is used to signify the entire system. 

\begin{figure}[htbp]
    \centering
    \includegraphics[scale=0.5]{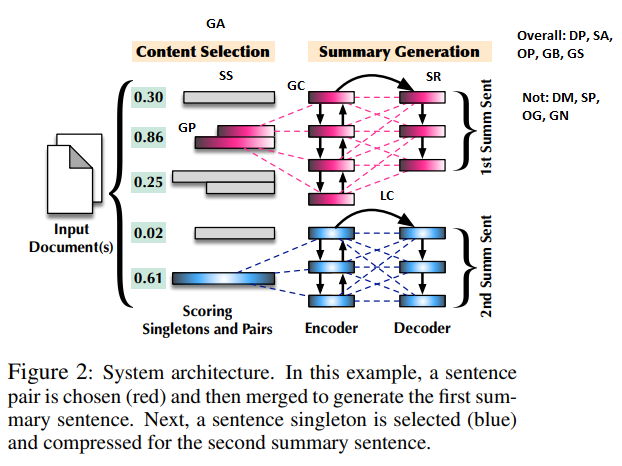}
    \caption{An example of a contribution-oriented schematic [Lebanoff et al, ACL 2019], with labels from the framework overlaid}
    \label{fig:lebanoff}
\end{figure}

\subsection{Example prioritising a schematic of the contribution: Fig. \ref{fig:lebanoff}}

Fig. \ref{fig:lebanoff} can be thoroughly described using the Richards-Engelhardt framework, which we will briefly do here without going into minutiae sub-figure discussion. 
There is grouping by alignment, by colour and by position. It uses two different linking graphical components (dotted lines and arrows). The content is quite schematic, with omission of details not important to the core contribution, such as layer labels and operations, as seen in the previous diagrams. Moody et al. \cite{moody2010visual} would describe this figure as demonstrating good \textbf{"graphic economy"}, whilst in Gestalt theory this would be "Prägnanz" \cite{Wertheimer1923UntersuchungenII}, language that the Richards-Engelhardt framework appears to lack. The framework does not facilitate discussion of the schematic and relatively minimalist content style, as the mode of depiction has a physical definition. Currently, the framework has the mode of depicting being schematic or realistic/precise, but this does not support the examination of individual graphic objects which is necessary for comparing between diagrams (see Fig \ref{fig:vectors}).

There are some important features that the framework does not currently include, such as the bracket object being used for grouping, which might be termed \emph{Grouping by Object}. The visual encoding here differs from other examples, in that the set of graphic components is taken from a smaller pallet: This is a simpler diagram by design. Unusual styling is present, in the graded colouring within the rectangles (it is not clear if this serves an encoding purpose). The rotated text also is visually notable, and may require head-tilting which physically changes the readers interaction with the diagram. The use of an icon to accompany "Input Document(s)" is a dual-coding unexpected given the schematic nature of the rest of the diagram. Note that the system inputs are data rather than physical pieces of paper, so the mode of correspondence remains non-literal. Moody \cite{moody2009physics} would describe this dual-coding as \textbf{"semiotically unclear"}, as there is not a 1:1 mapping between semantic constructs and graphical symbols. The precision of the diagram is also arbitrary; the numbers are indicative rather than meaningful, as are the number of rectangles. The numbers allow inference that the highest value is selected. The dotted lines suggest the network is Fully Connected, though there is one connection omitted top left to bottom right in the red network, perhaps accidentally.
We find Emphasis by Colour is used as well as Grouping by Colour. More generally, even within a simple example such as a pie chart, colour, spatial separation, and orientation or layout techniques can be used for emphasis. 
Semiotically, the object being signified is variable within this diagram, with sentence-specific numeric information alongside the system-level "document(s)". Indeed more broadly this diagram appears to be representing the "contribution" rather than the "system".

Where the authors reference the diagram in the text, it is described as an "illustration" \cite{lebanoff2019scoring}, perhaps reflecting that the mode of depiction is schematic rather than precise.

\section{Results and discussion}

The scholarly NLP system diagram domain is substantially more complex than the examples initially considered by Engelhardt and Richards, making intelligible implementation of the visual methods of their Fig. 2 challenging. This visual method has a table of visual encoding principles which was used to point at the instantiation of that principle. To clearly apply to complicated system diagrams following this visual method would seem to warrant an interactive digital visualisation rather than static diagram, due to the multiple applications of the same principles. In particular, unlike our domain, the examples the framework was tested on do not have extensive "subsets of graphical components" (sub-figures), and often employ a fairly limited number of visual encoding principles.

\subsection{Strengths of applying the framework}
We found it extremely useful to have the vocabulary and breadth of considerations provided by the framework, particularly around the graphical elements. It has successfully and usefully described how different principles are used to group and link, within diagrams in this domain. In general, we found the visual encoding principles unambiguous and straightforward to identify.

Applying this framework has also allowed us to discover internal inconsistency in some of the diagrams, and a number of errors in the creation of the diagrams appear to have been found. Perhaps authors and reviewers would benefit from examining diagrams with a critical eye, using Engelhardt and Richards Visual Encoding Principles.

\subsection{Refinements and extensions to make this a useful tool for reflection on (scholarly NLP) system diagrams}

Whilst, as the prior work of this paper attests, applying this framework was useful, there are a number of items which the framework requires for utility in our domain. It would be helpful to have standard terminology for some additional attributes of diagrams as part of the mode of depiction.

\paragraph{Author assumed knowledge and conventions} Assumptions made by the author, such as use of acronyms, mathematical notation, or context-specific conventions, are not reflected here. This may also usefully be considered as part of visual encoding in a general domain. We believe that, even in Engelhardt and Richards' own example (their Fig. 2) the assumed knowledge of Dutch language, and political acronyms, are important parts of the visual encoding. This may be significant enough to require calling out separately as part of "Depicting". 

\paragraph{Internal consistency}  For complex diagrams, "consistency" may be important. This does not only apply to how visual encoding principles are used, but also in natural language, consistency of capitalisation or use of symbols (e.g. Fig. \ref{fig:layercentric} uses ellipsis sometimes for arbitrary dimension, some lower case first letters, and even the "+" symbol is inconsistent between the caption and the diagram). Regardless of whether internal inconsistency is cognitively problematic, it is necessary to capture this in order to describe the diagram, and would provide an entry point to discussing differences in representational choice within the diagram.

\paragraph{Language} The choice of communication modes such as mathematical symbols, code, abbreviations, "iconic" symbols, or natural language is significant to the encoding. Whilst these can be considered as a subset of graphic components it perhaps does not do justice to the cognitive difference using these makes \cite{scanlan1989structured,boon2009models}. This could be structured, and consideration made of tone, language etc. This is a specific case of "Author assumed knowledge and conventions", but also is a different modality with different perceptual and semiotic considerations, and therefore seems worthy of specific attention.

\paragraph{Semiotics of examples vs systems} As part of the future work, it would be good to include semiotic or representational considerations (keeping within the framework's scope of "meaning represented in a diagram"). For example, if a systems diagram includes an example, as in Fig. \ref{fig:chess}, this shifts the diagrammatic representamen to being a specific instantiation of the system.


\paragraph{Common domain-specific symbols} Graphic components: within either "mode of depicting" or "visual encoding: depicting", it would be useful to have graphical components categorised to describe common symbol choices when comparing diagrams. Further, it is noted that for the visual encoding principles, quantitative, ordinal and nominal attributes are covered. The "depicting: picturing" can be extended to include other properties of the object that are in a general sense iconic, which would allow us to describe (for example) the differences in vector representations shown in Fig. \ref{fig:vectors}.

\paragraph{Sub-figures} In our domain, sub-figures (such as the layers) sometimes contain different representational choices to the "macroscopic" diagram. Engelhardt and Richards consider this (p204), and consider them to be nested visual structures, which we have been abbreviating as sub-figures. This indicates the utility of this framework at different levels of granularity. For discussion of complex diagrams, it might be suitable to add an external axes to indicate position of sub-figures and features, rather than overlaying this information onto the diagram. We would welcome discussion on other ideas for this.

\paragraph{Schematics} The framework does not capture the schematic (or otherwise) nature of the diagram, making comparison or discussion on "How much content is included or omitted" difficult. This applies not only in the mode of depiction, but also as part of the mode of visual encoding, as demonstrated by the wide range of graphical elements representing vectors in Fig. \ref{fig:vectors}. 

\paragraph{Physical emphasis} For non-physical systems, dimensions of discussion are reduced to being "non-literal" correspondence with "realistic/precise" depiction. It would be nice to have more language to help describe the use of non-physical metaphor and models. This could be part of the content work. 

\paragraph{Quantitative analysis} Allowing for quantitative discussion of diagrams at scale is challenging, particularly for complicated diagrams. This capability would be useful in order to do justice to the heterogeneity of visual encoding techniques employed, and allow for quantitative comparison (say, between different domains or media). This would seem to be a core benefit of covering the whole design space of visualisations, and as  such an extension to the framework to support quantitative analysis of visual encoding principles would be helpful.

\paragraph{Grouping by Object} In encoding by visual appearance, we found graphical components typified by "\{", and natural language labels, being used to group other graphical components. Whilst it could be argued this is a grouping by proximity or boundary, this does not capture that the visual object "\{" is used with spatial grouping to visually encode "nesting" at different levels of grouping. The right brackets of Fig. \ref{fig:lebanoff} is one such example of "\{" usage. In this example, it is to group together individual rows by "Input Document(s)", "1st Summ Sent" or "2nd Summ Sent", where spatial grouping is used to link the coloured rectangular objects with the descriptions. The "1st Summ Sent" grouping would not be clearly only applied to the last three coloured rectangles without the mediating "\{" symbol. As such, we claim that Grouping by Object is as different from other grouping mechanisms as a sheepdog is from a fence.  Note that the "\{" symbol should not be considered a "linking" symbol, as an arrow might be, as it applies a common property to a set of other objects and is therefore about "category", which Engelhardt and Richards have defined to be the goal of grouping (p206). (This "type" of grouping is often applied only after other visual encoding principles have been applied. It may be this is because it does not leverage Gestalt to the extent of the other Grouping principles.) 

\paragraph{Emphasis Principle} Additionally we propose "Emphasis" as a worthwhile extension to the visual encoding principles. It is most similar to the principle "Ordering", but fulfils a different function: 
\begin{itemize}
    \item[\textbullet] \textbf{Emphasising} is used to make visually salient particular aspects. \textit{Emphasising} answers questions such as \textit{What is most important?} and \textit{What should the reader look at first?}
    \begin{itemize}
        \item[\textendash] \textbf{Emphasis by colour} is a visual encoding principle often utilising bright or high contrast colours.
        \item[\textendash] \textbf{Emphasis by position} is done by placing visual primacy to certain elements, for example with a prominent position (extreme left, right or central), or providing lower spatial clustering with respect to other graphic elements or sub-figures.
        \item[\textendash] \textbf{Emphasis by uniqueness} is done by using a unique visual encoding principle for an element (or sub-figure). This could be a different boundary, colour, or graphical elements.
    \end{itemize}
\end{itemize}

\subsubsection{Ambiguities and other observations}

\begin{itemize}
\item[\textbullet] The scope of the framework, in terms of describing diagrams, was quite clear. We found it easy and clear to apply the visual encoding principles. However, it was unclear whether the caption should be considered as "part" of the diagram, and it would be good if that could be clarified as part of the future content work.
\item[\textbullet] We found that, even in our ecological setting, this worked well for standardised diagrams, and novel-but-thoughtful diagrams. The framework lends itself to diagrams that are consistently assembled. However, it lacks the vocabulary to describe errors or misleading visual encodings, which would be useful as part of validating this framework and providing feedback to diagram authors, a potential use case of this framework.
    \item[\textbullet] The vocabulary enables standard terminology to describe visual features of diagrams, and has been useful in describing figures in our domain. However, it does not yet provide guidance for semantic-content-focused vocabulary, and borderline topics such as graphic economy, and semiotics. The definition of "mode of depiction" also appears to need clarification in order to disambiguate precise/realistic and describe schematics and conventions.
    \item[\textbullet] The framework has facilitated description and discussion of complex systems diagrams, including sub-figures. However, the framework is not yet optimised for sub-figures, or for describing the layout of the diagram. Sometimes we have to infer the intention or meaning of the author, and clarification is only possible outside of the diagram (e.g. in text or speech).
    \item[\textbullet] For "positioning along an axis", we have assumed this axis does not need to be explicitly drawn, though for the infographics domain perhaps this distinction is useful. Either way, the framework allows for useful discussion on this topic. More generally, we felt that more precise definitions would be helpful for meaningful and unambiguous application of the framework.
    \item[\textbullet] In our domain, Authors sometimes seem to use visual encoding mechanisms in unconventional ways (cf. colour gradient in Fig. \ref{fig:datacentric}). This is perhaps not unusual in diagrams more broadly, and the framework would benefit from terminology for this.
    \item[\textbullet] Additional research that could be linked to this framework includes visual ontologies, which aim to completely describe objects and the relations between them. Of particular relevance are those from image processing and machine vision (such as \cite{maillot2004towards}), which have the potential to automate the application of this framework.
\end{itemize}


\subsection{Discussion on scholarly NLP system diagrams}
There is an overwhelming absence of relevant, established representations or visual notations for scholarly systems diagrams: As background for this study, we randomly sampled over 150 diagrams from top NLP conference proceedings. None utilised standards of UML or Block diagrams, nor used a diagrammatic output from the software. These established diagrammatic frameworks could, in principle, be used in the domain. We hypothesise that they are not due to their inability to express the new representational challenges of neural networks, such as the dimensional change, embedding, layers, etc.

There is an as yet un-examined semiotic difference between a system instantiated around an example, and one describing a general system. The former describes a single operation, rather than a system, though the meaning communicated may or may not be similar. Most diagrams we found represent the system, but some represent data pre-processing, data itself, or an example.

NLP systems can be created through visual programming languages, including specialist interfaces such as TensorFlow \cite{tensorflow2015-whitepaper} which has the TensorBoard visualisation tool for automatically creating a system diagram. We did not find any cases of this being used in conference proceedings. We hypothesise this is because a large proportion of research is done using other tools, and also because the level of granularity is too low. It may be that an additional, higher-level view on TensorBoard would facilitate system architecture diagrams that would be suitable for immediate export into conference proceedings. Repeatability, the ability to reproduce work, is a problem in Computer Science \cite{collberg2016repeatability} and better diagrams could be part of improving system transparency and lead to better repeatability and reproducibility.

We found that, even in very visually and semantically different diagrams, almost every Richards-Engelhardt grouping and ordering technique was deployed. This makes intuitive sense, as descriptions of software systems are principally about grouping and ordering. The prevalence of "Repetition", despite the use of varying graphic components for vectors and indeed the indeterminate nature of each one of the repetition uses, was unexpected. In conjunction with repetition, some diagrams used the ellipsis symbol "\ldots" to indicate continuity.

\subsection{Limitations and future work}
Using this framework, there are numerous avenues to investigate, including describing and comparing diagrams. These observations are in addition to the potentials described by Engelhardt and Richards. 

Other system diagrams, such as engineering drawings or those deployed in the discipline of Systems Theory, may benefit from examination with this framework. Additionally, there is the potential to investigate other ecological or "organically created" diagrams, such as biology school textbook diagrams or diagrams used in journalism, and to examine temporal or evolutionary differences in diagrams occurring in different domains. 

It would be useful to investigate perceptual properties of different visual encoding principles. Specifically for NLP system diagrams, a useful example to study would be repetition, including whether an ellipsis or a different number of repeated symbols has an impact on interpretability. Designing an experiment that respects the need to be distinct from any particular task would make this more challenging. Another relevant topic for our domain would be the perceptual and cognitive impact of simultaneous application of multiple visual encodings for grouping.

Further, contrasting the Richard-Engelhardt framework with domain specific tools, such as Physics of Notations for Software Engineering, might provide justification or insight into potential changes to those domain-specific guidelines.

\section{Conclusion}
To our knowledge, this is the first work to consider scholarly diagrams, as well as Natural Language Processing system diagrams, and amongst the first to apply the Richards-Engelhardt framework. We have successfully applied the framework in order to describe and discuss Natural Language Processing systems diagrams, identifying unconventional visual encoding choices and inconsistencies in the diagrams.

We have identified modifications and additions that would make the application of this framework more useful in our domain, and perhaps for systems diagrams more generally. We have also demonstrated that scholarly diagrams are an interesting avenue for investigation by the diagrammatic community.


\bibliographystyle{splncs04}
\bibliography{bib}
\end{document}